\documentclass[sigconf]{acmart}

\AtBeginDocument{%
  \providecommand\BibTeX{{%
    \normalfont B\kern-0.5em{\scshape i\kern-0.25em b}\kern-0.8em\TeX}}}

\copyrightyear{2023}
\acmYear{2023}
\setcopyright{rightsretained}
\acmConference[CHI EA '23]{Extended Abstracts of the 2023 CHI Conference on Human Factors in Computing Systems}{April 23--28, 2023}{Hamburg, Germany}
\acmBooktitle{Extended Abstracts of the 2023 CHI Conference on Human Factors in Computing Systems (CHI EA '23), April 23--28, 2023, Hamburg, Germany}\acmDOI{10.1145/3544549.3585871}
\acmISBN{978-1-4503-9422-2/23/04}

\begin{document}

\newcommand{\ryo}[1]{\textcolor{purple}{[Ryo: #1]}}

\newcommand{\system}{VR Haptics at Home}

\title[\system{}: Repurposing Everyday Objects and Environment \protect\\ for Casual and On-Demand VR Haptic Experiences]{\system{}: Repurposing Everyday Objects and Environment for Casual and On-Demand VR Haptic Experiences}


\author{Cathy Mengying Fang}
\email{catfang@media.mit.edu}
\affiliation{
  \institution{MIT Media Lab}
  \city{Cambridge}
  \country{USA}
}

\author{Ryo Suzuki}
\email{ryo.suzuki@ucalgary.ca}
\affiliation{
  \institution{University of Calgary}
  \city{Calgary}
  \country{Canada}
}

\author{Daniel Leithinger}
\email{daniel.leithinger@colorado.edu}
\affiliation{
  \institution{University of Colorado Boulder}
  \city{Boulder}
  \country{USA}
}


\begin{abstract}
This paper introduces VR Haptics at Home, a method of repurposing everyday objects in the home to provide casual and on-demand haptic experiences. Current VR haptic devices are often expensive, complex, and unreliable, which limits the opportunities for rich haptic experiences outside research labs. In contrast, we envision that, by repurposing everyday objects as passive haptics props, we can create engaging VR experiences for casual uses with minimal cost and setup. To explore and evaluate this idea, we conducted an in-the-wild study with eight participants, in which they used our proof-of-concept system to turn their surrounding objects such as chairs, tables, and pillows at their own homes into haptic props. The study results show that our method can be adapted to different homes and environments, enabling more engaging VR experiences without the need for complex setup process. Based on our findings, we propose a possible design space to showcase the potential for future investigation.
\end{abstract}


\begin{teaserfigure}
\centering
\includegraphics[width=\textwidth]{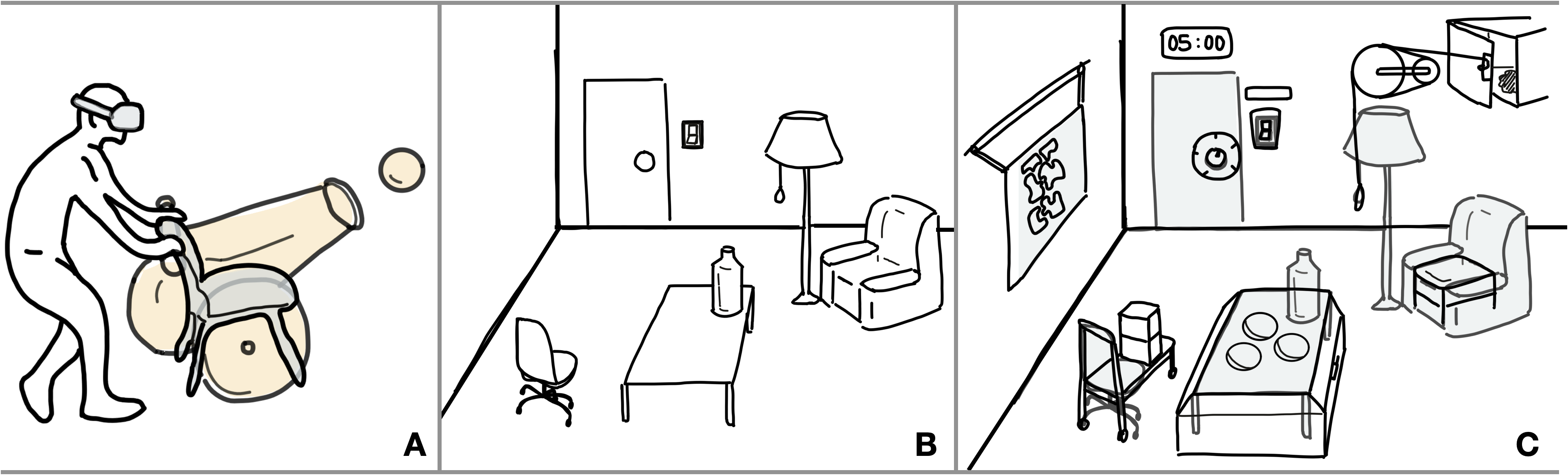}
\caption{We propose an idea that repurposes everyday objects (e.g., a chair) as passive haptic props (e.g., a cannon) to provide haptic feedback for virtual reality experiences (A). Figures B and C illustrate the potential scenarios of VR haptics at home --- by leveraging the functional affordances of various common indoor objects (B), users can transform an indoor space into an escape room (C) where they use complete tasks in virtual reality with physical objects.}
\label{fig:teaser}
\end{teaserfigure}


\maketitle

\section{Introduction}
Consumer virtual reality (VR) devices have shown great promise in creating visually immersive environments, and with devices like the Meta Quest~\footnote{\url{https://www.meta.com/quest/products/quest-2/}}, users are no longer tethered to the computer with wires and thus can play with VR wherever they wish. However, VR \textit{haptic} experiences are not the case for such casual and on-demand uses.
Today's haptic devices that go beyond simple vibrational feedback are not widely available for consumer VR users, as they are often costly, bulky, and complex, which significantly limits the opportunities for rich haptic experiences outside of research labs. 

What if, instead, we could turn our everyday environment into an adaptable, easy-to-use, and body-scale VR haptic environment?
In this paper, we explore VR Haptics at Home, an idea of repurposing everyday objects and environment to create casual, on-demand, yet engaging VR haptic experiences at home without the need for any special equipment.
The idea of leveraging passive haptic props itself is not new, as it has been explored in the literature (mostly known as Substitutional Reality~\cite{simeone2015substitutional}). However, most of the previous work has been studied in a controlled lab setting, leaving questions about how this approach can be adapted to various rooms in real-world environments.

To fill this gap, this paper contributes an \textbf{\textit{in-the-wild study}} conducted in each participant's home, rather than in a research lab, to better understand how such ideas could scale and how difficult the configuration process could be.
To this end, we developed a proof-of-concept prototype that allows users to configure their own haptic props on-demand through a simple setup, which allows us to investigate how this method can be adapted to different homes and environments.
In our study, we asked eight participants to use our system to turn their surrounding objects into haptic props for interactive VR games. More specifically, they used their own \textit{table as a ground} for a whack-a-mole game, \textit{chair as a canon} for a shooting game, and \textit{pillow as a cat} for petting.
The study results show that our method can be adapted to objects in different homes, and the extra steps to configure the objects are considered to be minimal and easy to follow. In addition, while some tasks were more negatively affected by imprecision in hand-tracking than others, overall, our method can effectively provide haptic feedback and enable more engaging VR experiences.
Through the study and iterative design explorations, we also learned that the concept of VR haptics at home can go beyond these three object modalities and there are a lot of untapped opportunities for this idea. For example, we found that not only the shape but also the affordances of objects, such as compliance, mechanism, and texture, could enhance the rich tactile feedback for passive haptics. Moreover, we can reuse a part of the object, rather than focusing on the entire form, to broaden the adaptability and generalizability of such haptic props.
To showcase such potential, we discuss a possible design space with accompanying exemplary scenarios, which we hope could inspire the research community for further investigation.

\section{Related Work}

\subsection{Passive Haptics}
Passive haptics is a technique that repurposes physical props or environments to create haptic sensations in VR and AR~\cite{insko2001passive, hoffman1998physically, daiber2021everyday}. Passive haptic devices allow users to control and manipulate 3D virtual models with greater flexibility and accuracy~\cite{hinckley1994passive, shapira2016tactilevr, zielasko2019passive, zhou2020gripmarks, henderson2008opportunistic}, without the need for special-purpose haptic devices, making them highly adaptable and deployable.
However, one challenge is the mismatch between virtual and physical objects~\cite{rock1964vision}. To address this problem, haptic retargeting~\cite{azmandian2016haptic, fang2021retargeted} is a technique that uses visual illusion to redirect the user’s hand when touching a virtual object. 
Similarly, other techniques, like Annexing Reality~\cite{hettiarachchi2016annexing} and Sparse Haptic Proxy~\cite{cheng2017sparse}, build on this approach by remapping the geometry of physical props and environments on-the-fly. 
Alternatively, encountered-type haptics~\cite{mcneely1993robotic} aim to provide dynamic passive haptics for whole-body haptic interaction (e.g., \textit{RoomShift}~\cite{suzuki2020roomshift}, \textit{MoveVR}~\cite{wang2020movevr}, \textit{ZoomWalls}~\cite{yixian2020zoomwalls}, \textit{CoVR}~\cite{bouzbib2020covr}). 
Instead of robotic systems, researchers have also explored human actuation to create dynamic haptic environments~\cite{cheng2015turkdeck,cheng2014haptic,cheng2017mutual}.
However, these active haptic systems are currently mostly limited to research lab use due to high costs, large size, and safety concerns.

\subsection{VR with Real-World Environment}
To address this problem, recent works have also started looking into how to further blend the virtual and physical environments for VR to provide more immersive experiences that go beyond mobile and hand-held haptic proxy.
Substitutional Reality~\cite{simeone2015substitutional, eckstein2019smart}, for example, reuses rather than replaces surrounding physical props as objects in the virtual space. 
Other systems use a mix of AR and VR to dynamically represent real-world objects and stimuli or reconstruct physical space ~\cite{yang2019dreamwalker, hartmann2019realitycheck,lindlbauer2018remixed,lin2020architect, tao2022integrating}.
Our work builds upon concepts presented in substitutional reality, but we contribute to the results and insights from the in-the-wild study to evaluate the idea in the real-world environment rather than in a controlled lab setting. We also contribute to the design space to expand the range of everyday objects that can be incorporated into VR experiences.

\begin{figure*}[t!]
\includegraphics[width=\textwidth]{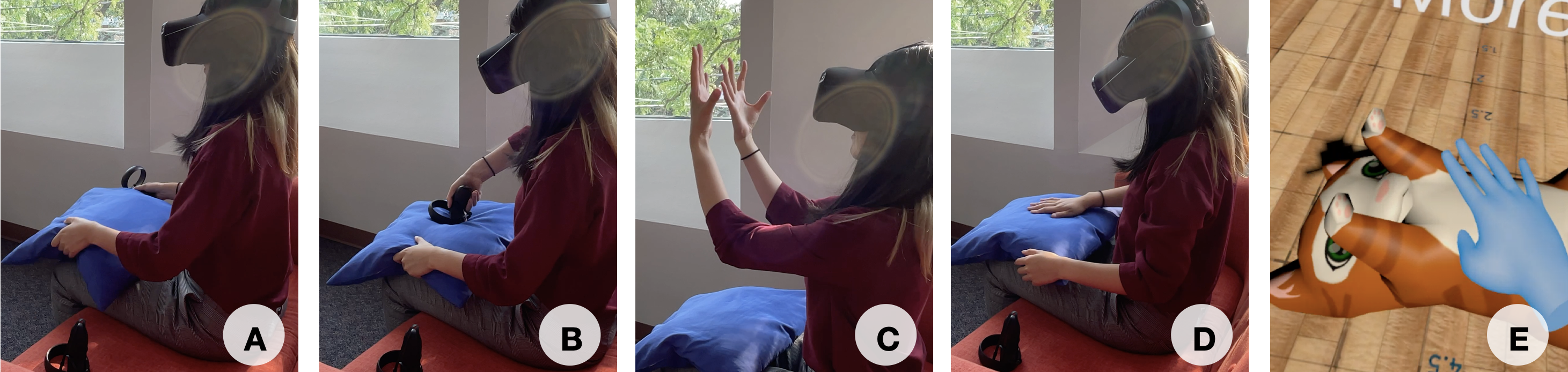}
\caption{(A) The user places the object of interest at a desirable location. (B) The user aligns the controller against the object to place the virtual element. (C) The user places the controllers aside somewhere safe and looks at their hand to switch to hand-tracking mode. (D) The user feels the haptic feedback from the physical object while in VR (E) sees the hand touch a virtual cat.}
\label{fig:setup-process}
\end{figure*}

\section{VR Haptics at Home: Proof-of-Concept System}
\begin{figure}[b!]
    \centering
    \includegraphics[width=\linewidth]{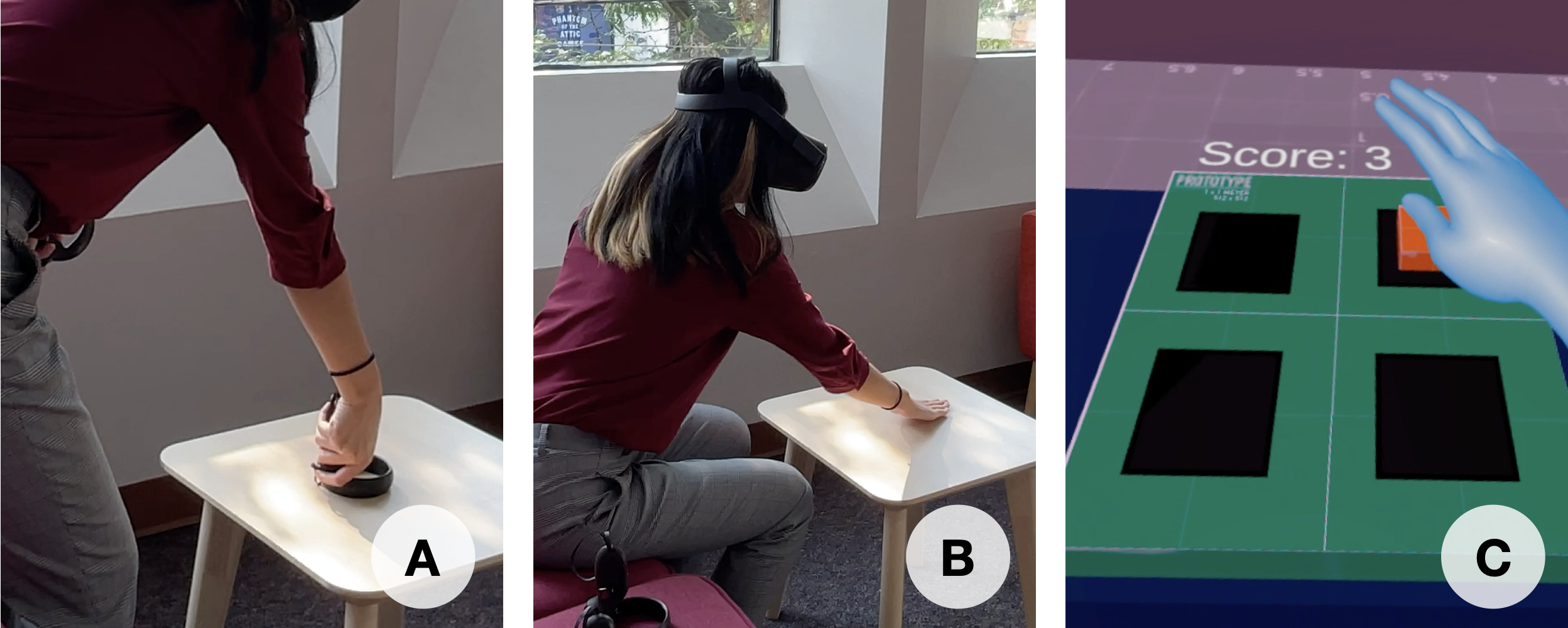}
    \caption{The user aligns the controller against a table surface to place the game platform (A). They can tap on the table and receive haptic feedback (B) as they hit the virtual moles in the Whack-a-Mole game (C).}
    \label{fig:demo-table}
\end{figure}

To investigate the VR Haptics at Home concept, we developed a proof-of-concept prototype with the Unity 3D engine. 
Our system consists of three interactive VR games: 1) pet a cat, 2) whack-a-mole, and 3) shoot monsters (See Figure \ref{fig:setup-process}, ~\ref{fig:demo-table}, and~\ref{fig:demo-chair} respectively).
We deployed our program on the Meta Quest and used Meta's hand tracking API (1.40). We found the accuracy of hand tracking to be sufficient to create a proof-of-concept. We note that our implementation is not limited to Meta's tracking technology. To repurpose physical objects, the user first prepares the object to be within reach (Figure \ref{fig:setup-process}A), and in the headset draws the guardian boundary to include the object. They place the controller against the physical object and press the trigger to place the virtual interactable object (Figure \ref{fig:setup-process}B). Once all the virtual objects have been placed, users then place the controllers face down somewhere safe and allow the system to automatically switch to the hand-tracking mode (Figure \ref{fig:setup-process}C). Lastly, the user can directly interact with the virtual object with their hands (Figure \ref{fig:setup-process}D). We have also considered leveraging hand tracking to, e.g., pinch to place the virtual interactable object, but to ensure reliable performance during the study, we used controllers as the reliable setup method.

\section{In-the-Wild User Study}
We tested with 8 participants (male: 5; female: 3; mean age 21.4; all with VR experience) and paid each a \$15 gift card. Participants used their own Meta Quest (first or second generation) and performed the tasks remotely at their homes or offices. 
The three scenarios created for the study include “Whack-a-mole” where the participants hit the mole 10 times, “Pet a cat” where the participants rub the cat’s belly 10 times, and “Shoot monsters”, where the participants aim and hit 20 monsters. There were two conditions: the {\it control, no-haptics condition} which every task was performed in mid-air, and the {\it haptic condition} where participants followed instructions in VR to configure the physical objects. The order of the scenarios and conditions was counterbalanced. Following the instructions in a Google Doc, participants set up the Quest headset for hand tracking and drew the guardian boundary in a space where "there is a clean tabletop surface, a moveable chair, and a small cushion or pillow" (the exact instruction given). After each task, they took off the headset and filled out a seven-question Likert-scale questionnaire about the task in a Google Form. Finally, the investigator conducted a semi-structured interview via voice call where participants elaborated on their preferences and experiences during the setup process and interactive scenarios.

\section{Results and Discussion}
Combining the quantitative results and feedback from post-study interviews informed the following insights and design recommendations. We performed a two-way ANOVA on the quantitative data where an align-ranked transform was performed as the scores of the questions do not have a normal distribution (Shapiro-Wilk normality test p<.001).

\begin{figure}[b!]
    \centering
    \includegraphics[width=\linewidth]{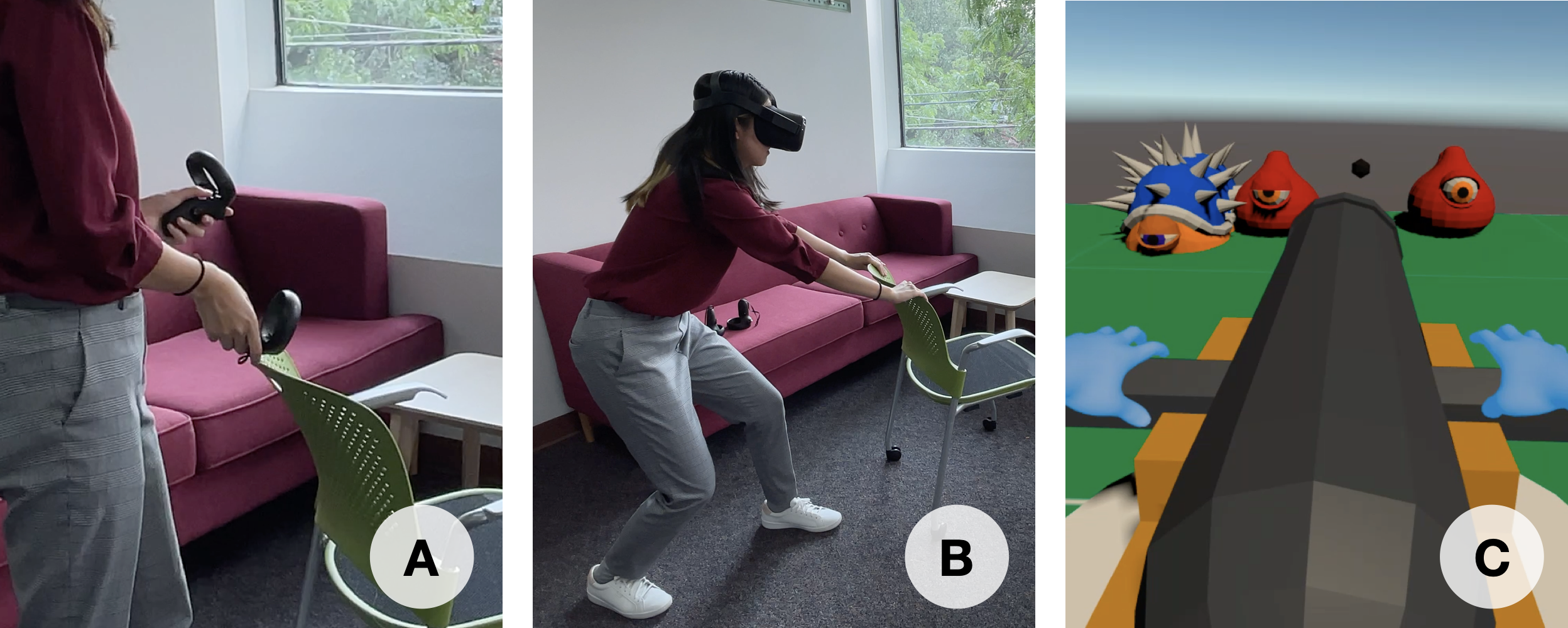}
    \caption{The user aligns the controller against the back of a chair to place the cannon (A). They can rest their hands on the chair, move it around and receive haptic feedback (B) as they rotate the virtual cannon to aim at the monsters (C).}
    \label{fig:demo-chair}
\end{figure}
\subsection{Haptic Feedback}
To understand whether the repurposed object indeed provided appropriate haptic feedback, we asked the following questions. Q1: The virtual object felt real when I touched it. Q2: The virtual object felt like it was there when I touched it. (1–strongly disagree, 7–strongly agree). For both questions, the haptic condition was rated significantly (p<.001) higher than the no-haptic condition (Q1 Haptic: M=3.7, SD=1.6; No-haptic: M=1.6, SD=1.2; Q2 Haptic: M=4.4, SD=1.8; No-haptic: M=2.1, SD=1.9). Evidently, the temporal coincidence of the hands interacting with physical objects helps reaffirm the virtual objects' realness. Most participants reported having found haptic feedback to be helpful, providing additional cues and feedback for their actions. For “whack-a-mole”, the surface of the table provides a stop to the hand’s downward motion when “mole” is hit, and participants reported that “slamming down on a table felt satisfying” (P5). Similarly, the back of the chair provided a cue for “where to rest the hands” (P4) on the virtual cannon. Participants also commented specifically on the similarity between the texture and compliance of the pillow and the cat's fur. Petting a “cushion” made the experience more immersive; participants were “surprised by how much more real it felt” (P8). There were variations in how well the haptic feedback matched the expectation and visual feedback. Both the questionnaire results and post-study interview reveal that participants preferred “Pet a cat” the most, followed by “Whack-a-mole” and “Shoot monsters”. The texture and compliance of a cushion coarsely match with those of a furry animal, thus the haptic feedback still positively contributes to the experience despite the difference in the shape of the cushion and the cat. Even though the table surface is able to cue where the hand stops for “Whack-a-mole”, the binary state of the button is not produced with a flat table. The virtual buttons "had no compliance" and felt more like "holograms" (P6). Finally, while the swivel mechanism of a chair affords the rotating motion of the cannon, P7 noted that “the cannon was too easy to spin when it appeared heavier”. To mitigate this, designers could use techniques like haptic retargeting \cite{azmandian2016haptic} or make the cannon appear lighter visually. To summarize, when considering an object or part of an object to repurpose as a haptic prop, designers should choose everyday objects that can complete the feedback loop of users' actions and use visual-audio cues to fill in the gaps.

\begin{figure}[t!]
        \centering
        \includegraphics[width=\linewidth]{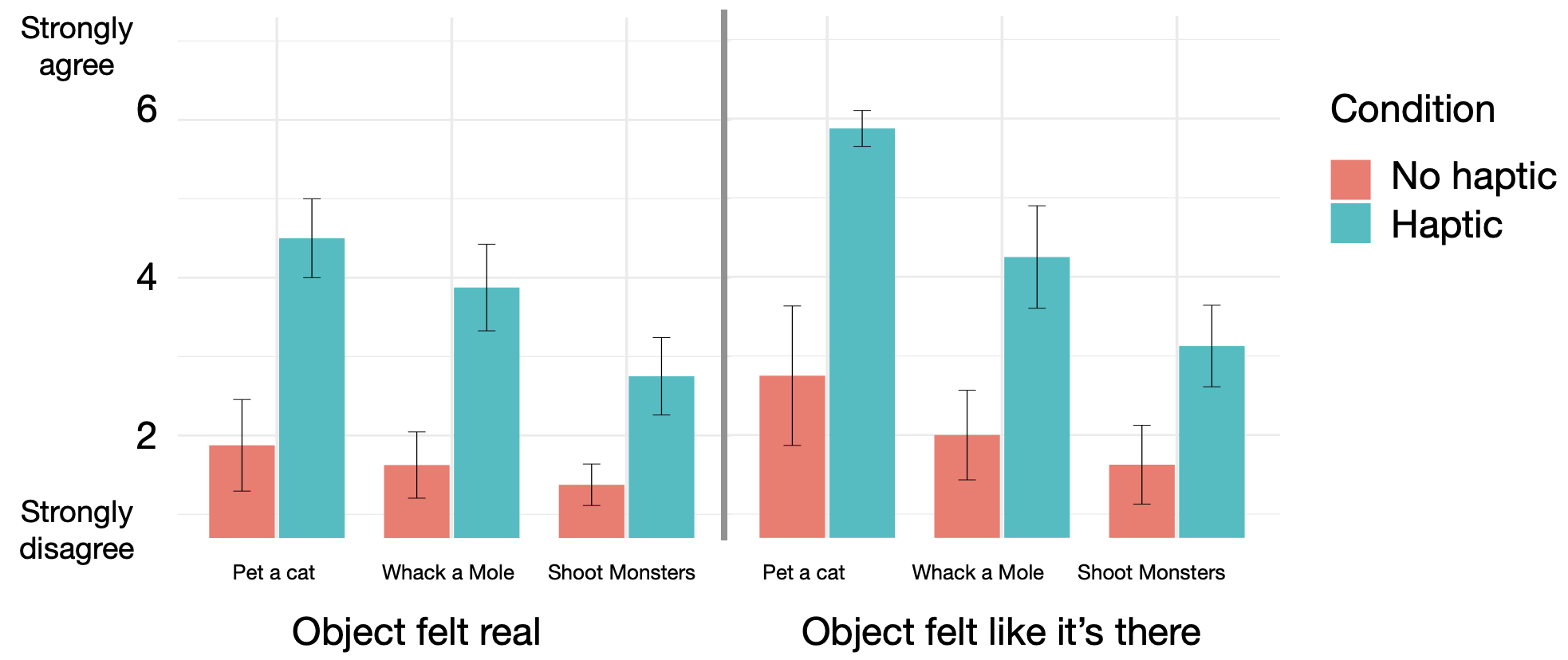}
        \caption{Results of haptic feedback. Error bars: SE.}
        \label{fig:graph-1}
\end{figure}

\begin{figure}[b!]
   
        \centering
        \includegraphics[width=\linewidth]{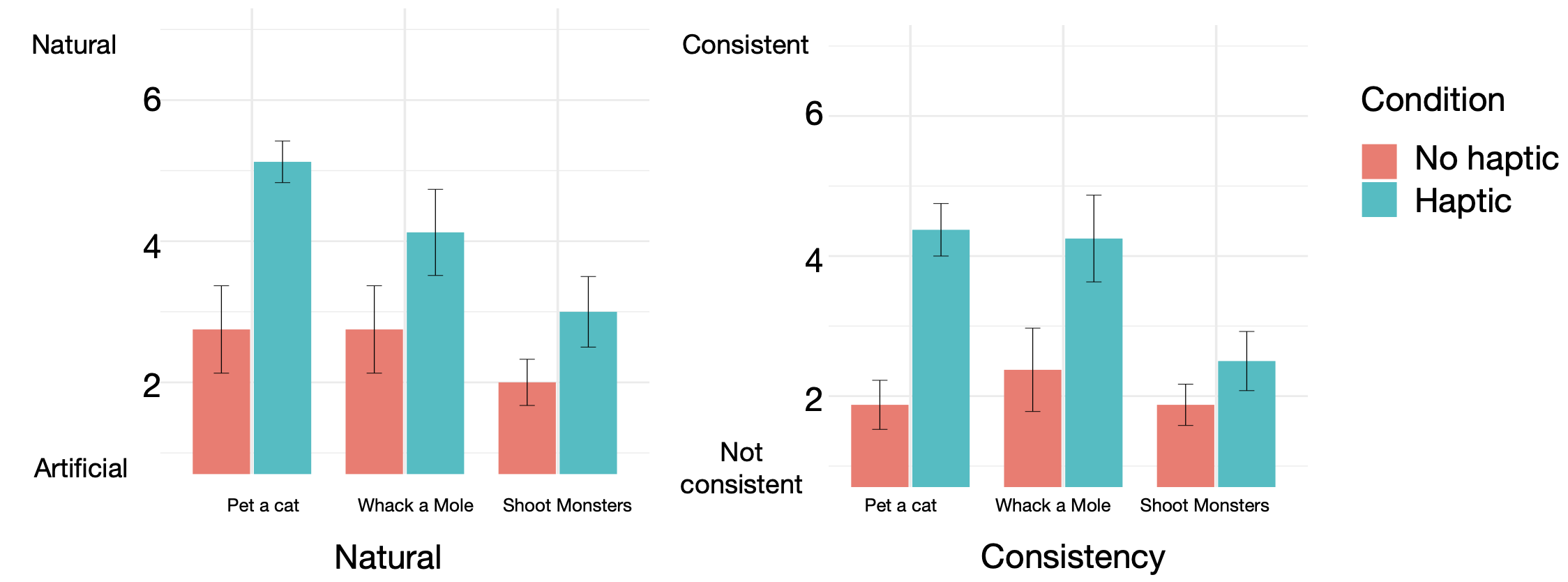}
         \caption{Left: Results of naturalness. Right: Results of consistency. Error bars: SE.}
        \label{fig:graph-3}
\end{figure}

\subsection{Set Up and Configuration Process} \label{setup}
We followed the original 1–7 scale of the NASA TLX questionnaire \cite{hart1988development}, where a lower score means less workload and more desirable. The mean score for the haptic condition was 4.02 (SD=1.11), which is higher than that of the no-haptic condition (M=2.82, SD=1.82). This was expected, as the haptic condition required extra steps that are not part of a typical VR experience and hence require more workload. However, participants responded positively about the extra steps. In post-study interviews, participants reported that the setup process was "easy" (P2) and "the instructions were clear" (P4). They were able to "find the objects easily" (P1) because they felt "familiar with the room spatially" (P1), and they appreciate that the "pillow was adjustable" (P5). Though some participants expressed concerns regarding hitting the wall and damaging the controller. Some commented that including walls and large objects within the guardian boundary is counter-intuitive as typical VR experiences happen in an obstacle-free space. Participants responded well to soft mobile objects such as a pillow. These objects are lightweight thus easy to move, and they are soft so they cannot cause any harm during setup or interaction. Rigid chairs are accessible but require more effort to move. P1 commented that the chair had a "huge presence...and took physical effort to move around". In summary, we recommend designers take advantage of mobile objects that are soft and lightweight, such as books, boxes, and cushions. For static objects, we recommend indicating their presence to prevent the users from running into them. Given the current limitation of object tracking and to reduce the burden on the users, we recommend using a single physical object as a haptic prop throughout the gameplay.

\subsection{Controller and Hand Tracking}
To understand the effect of the input method (i.e., controller and hand tracking), we adapted the following two questions from \cite{witmer1998measuring}. Q3: How much did the controller tracking/hand tracking interfere with the performance of assigned tasks or with other activities? (1-not at all, 7-interfered) Q4: How well could you move or manipulate objects in the virtual environment? (1-not at all, 7-extensively) Although there was no statistically significant effect of tracking interference, "Shoot monsters" was rated higher (more inference) for both conditions. From participant feedback, having physical props allowed them to "know where to stop the hand" (P5) and can e.g., "pet [the cat] very naturally" (P8). They reported that the overall hand tracking worked well except for the "Shooting monsters" task, where the grasping gesture sometimes was not detected accurately.

\begin{figure}[t!]
    \centering
    \includegraphics[width=\linewidth]{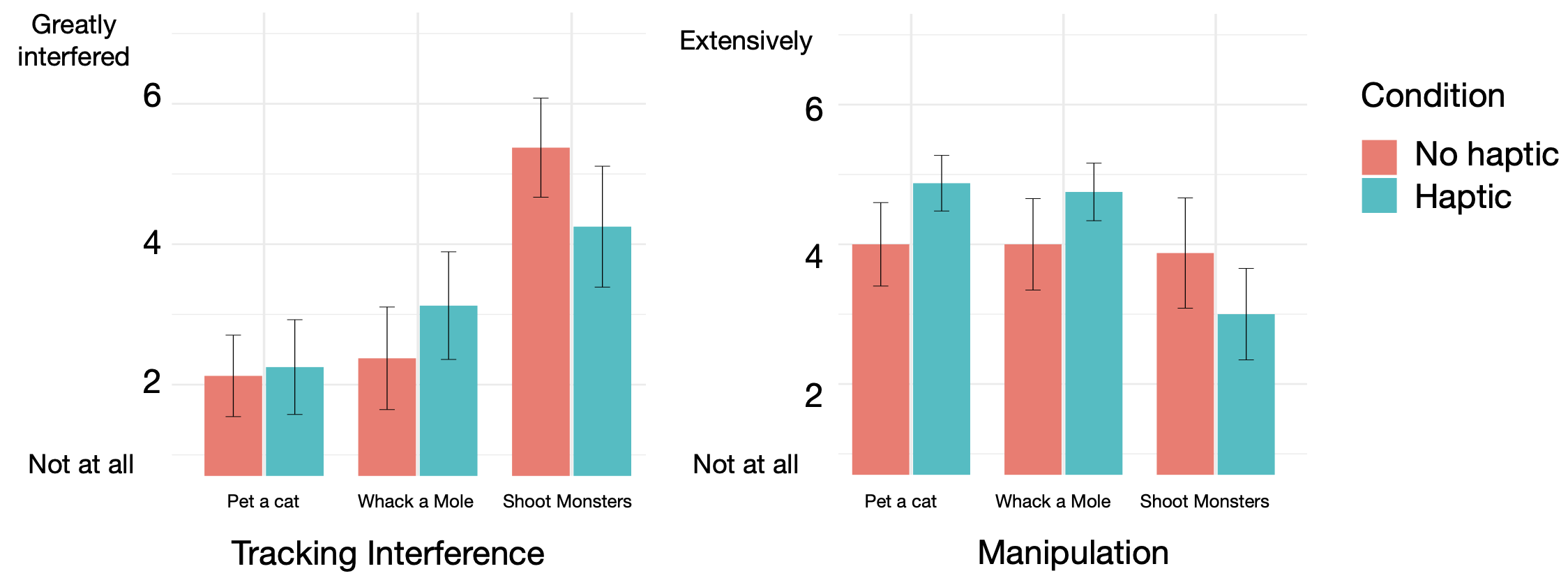}
    \caption{Left: Results of tracking interference. Right: Results of object manipulation. Error bars: SE.}
    \label{fig:graph-2}
\end{figure}

\begin{figure} [b!]
        \centering
        \includegraphics[width=\linewidth]{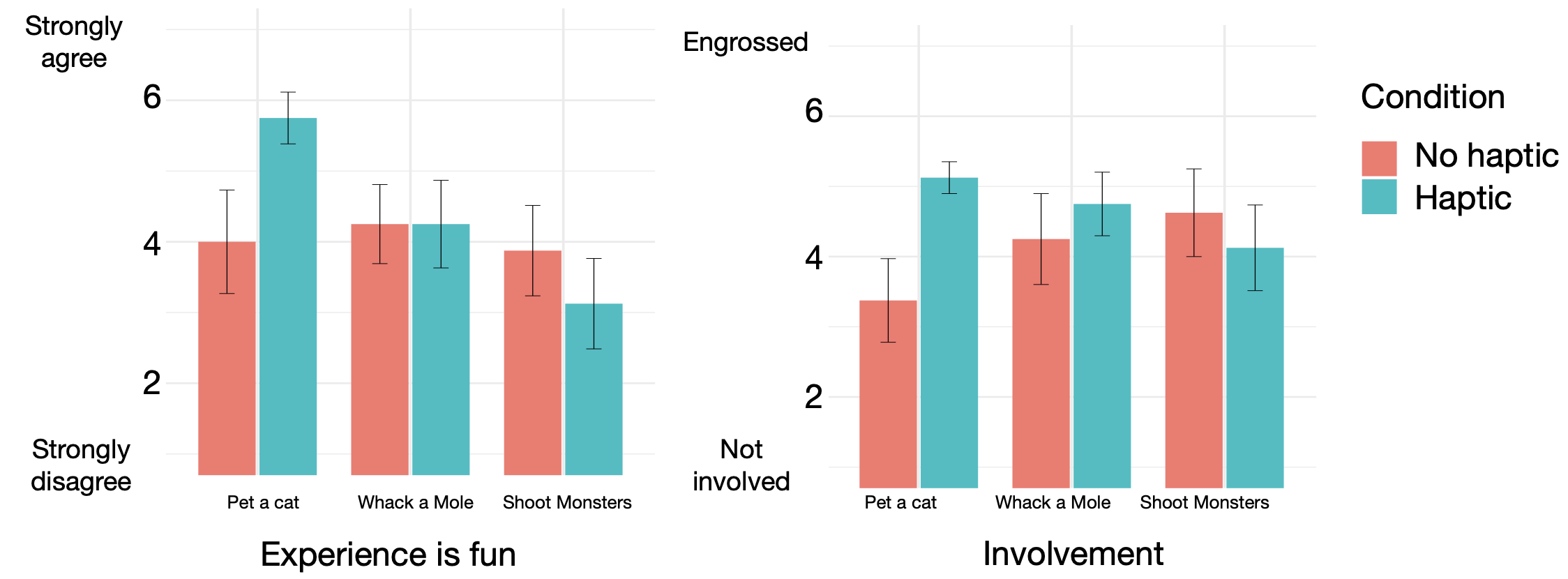}
        \caption{Left: Results of funness. Right: Results of involvement. Error bars: SE.}
        \label{fig:graph-4}
\end{figure}

\subsection{Immersion, Realism, and Fun}
To understand participants' experience, especially their sense of presence, in the virtual environment while engaging with real-world objects, we asked the following questions: Q5: How natural did your interactions with the environment seem? (1-artificial, 7-natural) Q6: How much did your experiences in the virtual environment seem consistent with your real-world experiences? (1-not consistent, 7-consistent) Q7: This task was fun. (1-strongly disagree, 7-strongly agree) Q8: How involved were you in the virtual environment experience? (1-not involved, 7-engrossed). Q5, Q6, and Q8 are from the Presence Questionnaire \cite{witmer1998measuring}. The haptic condition was rated as more natural (M=4.1, SD=1.6, p<.001) than the no-haptic condition (M=2.5, SD=1.5), and "Pet a cat" was rated as the most natural. The haptic condition was rated as statistically more consistent (M=3.7, SD=1.6, p<.001) than the no-haptic condition (M=2.0, SD=1.2), and likewise "Pet a cat" was rated as the most consistent. In the follow-up interview, most participants noted the order of the most to least realistic task is "Pet a cat", "Whack-a-mole", and "Shoot monsters". There was a novelty effect for "Shooting monsters" as some commented they "never projected a cannon on a chair before" (P1) but found it nevertheless an interesting experience. Thus, "Shoot monsters" is the least close to a real-world experience. Although there was no statistical significance for fun and involvement during post-study interviews, participants specifically commented on how they "didn't expect [petting] the cat to be so fun" (P6) with a pillow as a substitute, and a couple of participants reported “time went by faster; [they] didn’t notice the progress bar” (P8).


\begin{figure*}[t!]
\includegraphics[width=\textwidth]{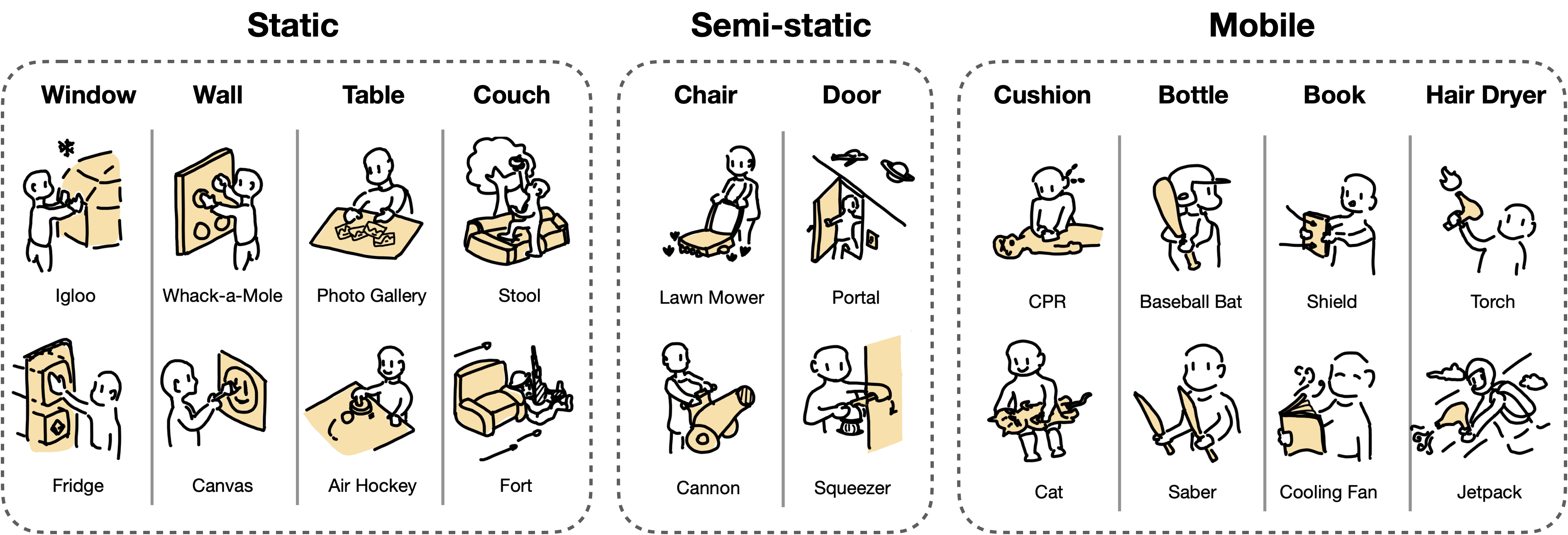}
\caption{The design space of our idea. We illustrated two scenarios for each object to show the rich design opportunity.}
\label{fig:design-space}
\end{figure*}

\section{Design Space}
Through the participants' feedback and iterative design process, we also learned that the idea of VR haptics at home can go beyond the three demonstrations we developed. Therefore, based on our exploration, this section explores the possible design space by illustrating exemplary VR experiences for future investigation (Figure \ref{fig:design-space}). 
Moreover, we discuss how the \textit{affordance} of physical objects, rather than similarities in shapes, can affect virtual haptic experiences.


\subsection{Mobility}
For VR haptic scenarios, mobility depends on the object’s weight, size, and relation to the indoor environment, which provides constraints to the possible interactions. 

\subsubsection{Static Objects and Surfaces}
Static objects or surfaces are large and structural and thus are most suitable for setting reliable boundaries. 
Going beyond repurposing the entirety of the object, we can focus on the flat surface as an affordance. For example, tangible interactions such as organizing a digital photo gallery where the user swipes and moves the photos presented on a surface. We can also simulate haptic interactions such as playing the piano where the hand encounters a surface when touching the virtual interactables. Flat surfaces can be a constraint for mobile objects. A table air hockey striker can only move along the plane of an air hockey table.

\subsubsection{Semi-Static Objects}
We define semi-static objects as heavy objects that can be moved within certain constraints. Beyond being used for sitting, a chair can also be pushed along the floor. We can leverage the constrained motion and repurpose the chair as a lawnmower, a similarly heavy object that one does not pick up and only moves across the ground. If the chair swivels, the constrained rotating motion allows the chair to be repurposed as a game cannon. Likewise, a door naturally serves as a portal to a different world, but we could also leverage constrained motion around the hinge.

\subsubsection{Mobile Objects}
These objects are lightweight and can afford interactions such as grasping and gripping along with swift movements due to their mobility. A water bottle that can be grabbed with the whole hand and moved around is suitable for replacing a saber. In addition, mobile objects can afford bimanual manipulations. Once again, a water bottle can be grabbed with both hands and acts as a baseball bat. A mobile object is not constrained by its location. A cushion placed on the ground can be repurposed as a CPR dummy, but when it’s placed on the lap, it can be replaced as a pet cat.

\subsection{Functional Affordance}
Furthermore, building on the findings in Substitutional Reality~\cite{simeone2015substitutional} where the similarity in affordance allows for a more believable substitution, we focus on leveraging the functional affordance of everyday objects. 

\subsubsection{Compliance}
The home environment conveniently provides compliant objects that are otherwise difficult to simulate using active haptic devices. A cushion is a convenient substitute for a small animal that allows the user to feel the softness when touching and petting the animal. In addition, compliant objects can absorb impact from large forces or fast actions. For example, a couch or a cushion could be used for CPR training where the user exerts a large downward force, or for a boxing game where the player hits the punching bag at a high speed. 

\subsubsection{Mechanism}
Owing to their purposes, objects or components in the room have functional mechanisms that can be repurposed in games and VR environments.
For example, rotating knobs and switches are common gameplay elements that players interact with to control the scene. 
Beyond using the mechanisms for their original purpose, we can focus on the constrained motion that they afford. A door handle affords a lever that can be repurposed as the handle of a lemon squeezer. The door affords a rotating motion constrained by the hinge that can only be moved along the edge of a circular plane at the height of the door handle. A drawer can only be pushed or pulled linearly.

\subsubsection{Temperature}
One may think about the thermal conductivity of a material as an affordance to design experiences that involve temperature. Touching glass windows generates a cooling sensation that can be repurposed for a cold environment like an igloo. On the other end of the spectrum, being wrapped around in a blanket over time feels warm and can simulate sitting near a fireplace. User-generated actions can also help create a breeze at different temperatures. Using a hair dryer can generate warm air and using a fan or fanning oneself with a book can generate a cool breeze.

\subsubsection{Texture}
Texture plays a key role in providing information about the material of an object, and at the same time, it is a challenging feature to recreate using active haptic devices. Existing objects in the room provide a rich library of textures. A fuzzy cushion is similar to an animal's fur, and a carpet is analogous to a grass field. With the help of visual cues, it is not difficult to imagine touching the curtain as feeling a willow tree or a giant's long hair.






\section{Limitation and Future Work}

\subsubsection*{Object Tracking and Recognition} 
To reduce the manual effort for configuration, we are interested in exploring real-time object tracking using LiDAR and object recognition. In this case, once an object has been located and configured using a controller, we assume that the object is either stationary or follows the hands’ movement. Being able to adapt to the physical objects properties and features would further enhance the immersivity. For example, being able to recognize the texture of a cushion and adapt the cat's fur to the texture or change the curvature of the cannon's handle to match that of the back of the chair.



\subsubsection*{Variability in Objects} Even though we chose common objects that one can find in a room, we expected variations in the objects in our participants’ space. 
For example, the back of the chair had variations in curvature and some participants’ chairs could not swivel, which limits how the virtual cannon could be manipulated. 
Also, from the participants’ feedback, the mismatch between the flat table surface and the binary state of the “mole” was prevalent and affected the experience more than other factors. 
Future work should investigate how to address these variations of physical objects through, for example, automated virtual shape deformation.


\subsubsection*{Fill the Gap of the Design Space} 
Finally, as discussed in the Design Space section, there is an interesting design challenge in leveraging \textit{functional affordances}, rather than the shape of an object itself. Moreover, our design space exploration is just an initial investigation, and we did not demonstrate or prototype most of the exemplary scenarios illustrated in Figure~\ref{fig:design-space}. Future work should further investigate and demonstrate other possible VR haptic applications by filling the gap in the current design space, and then examine how these affordances allow for richer haptic experiences in real-world environments. In addition, electric or motorized appliances such as an air-conditioner and Roomba as in MoveVR\cite{wang2020movevr}, could afford even more dynamic and diverse virtual applications.

\section{Conclusion}
We present VR Haptics at Home, where users repurpose common, existing objects and surfaces at home to provide casual, on-demand haptic feedback for more immersive VR experiences. We describe the rich design space using this method and developed a couple of applications. We conducted an in-the-wild study of our applications with VR users, from which we present design recommendations for future designers and developers that wish to use this technique.

\bibliographystyle{ACM-Reference-Format}

\bibliography{template}

\appendix

\section{Exemplary scenarios}
We selected and built three scenarios to demonstrate objects with various affordances.

\subsection{Whack-a-Mole} A Whack-a-Mole game is a game where the player hits the “moles” that peeks through the game platform before they disappear. In a typically VR experience, the user’s hand feels nothing when hitting a virtual button or the mole in mid-air, and the hand also goes through the virtual game platform. In this scenario, we leverage the flat surface of a tabletop to be a game platform. When the user presses the virtual button to initiate the game and taps the mole to win points, their hand is stopped by the tabletop surface.  For implementation, we attach the game platform to the front of the controller but the platform is hidden at first. When the user places the ring of the controller flush against the flat surface and presses the trigger, the game surface is enabled and placed where the controller ring is, which then aligns with the physical surface (Figure \ref{fig:demo-table}).

\subsection{Pet a Cat} In this VR experience, users interact with a virtual pet. When a user reaches out their hand and wants to pet the animal, their hand normally feels nothing. For this scenario, we repurpose a cushion as a cat. When the user pets the cat, they feel both the texture and the compliance of the cushion that mimics the haptic feedback from that of a cat’s belly. To implement this, similarly, we attach a virtual cat to controller. When the user aligns the controller against the center of a cushion and confirms with a trigger, the top of the cushion aligns with the cat’s belly. When the user pets the cat, they hear the cat cry and purr and also see it wiggle its body as a gesture of affinity (Figure \ref{fig:setup-process}E).

\subsection{Shoot Monsters} The user controls a cannon to aim and shoot the monsters to win points. As the monsters appear in different locations, they would need to grab the handle of the cannon to rotate and aim at the monsters. If the user controls the cannon with their hands instead of a joy stick, their hands would go through the handle and feel no haptic feedback that would give them a sense of control of the cannon. Here we use a chair and leverage the movement constraints to enhance this VR experience. The cannon appears to be heavy and cannot be easily picked up, and so is the chair. The cannon has wheels that allows it to move along the ground, and its handles signals that it can be rotated. Even though the shape of the chair does not match the virtual cannon, the back of the chair affords grabbing and the rotating motion of the moveable chair matches the same motion constraint of the cannon. For implementation, we press the controller against the top edge of the back of the chair, such that the handle of the cannon then aligns with the chair’s back. The virtual handles are grabbable objects that follow the users' hands, which allows the movement of the virtual cannon and the physical chair to be in sync (Figure \ref{fig:demo-chair}).


\end{document}